\documentclass[sigconf]{acmart}

\usepackage{enumitem}
\usepackage{cleveref}
\usepackage{xcolor, colortbl}
\usepackage{csquotes}
\usepackage{tabularx}
\usepackage{makecell}
\usepackage{multirow}
\usepackage{booktabs}
\usepackage{balance}

\usepackage{arydshln}
\newcommand*\dhline{\specialrule{0pt}{1pt}{0pt}\hdashline[.4pt/3pt]\specialrule{0pt}{0pt}{1pt}}

\usepackage[group-separator={,}]{siunitx}
\usepackage{fontawesome5}
\usepackage{tikz}
\usepackage{tikzscale}
\usetikzlibrary{arrows.meta}
\usetikzlibrary{shapes.symbols}
\usetikzlibrary{shapes}
\usetikzlibrary{decorations.pathreplacing}
\usetikzlibrary{backgrounds}
\usetikzlibrary{fit}

\usepackage{diagbox}

\usepackage{todonotes} %

\usepackage{tablefootnote}

\usepackage[absolute,showboxes]{textpos}

\usepackage[labelformat=simple]{subcaption}

\graphicspath{{figures/}}

\usepackage{url}

\usepackage{hyperref}

\copyrightyear{2025}
\acmYear{2025}
\setcopyright{cc}
\setcctype{by}
\acmConference[CCS '25]{Proceedings of the 2025 ACM SIGSAC Conference on
Computer and Communications Security}{October 13--17, 2025}{Taipei, Taiwan}
\acmBooktitle{Proceedings of the 2025 ACM SIGSAC Conference on Computer
and Communications Security (CCS '25), October 13--17, 2025, Taipei,
Taiwan}\acmDOI{10.1145/3719027.3765096}
\acmISBN{979-8-4007-1525-9/2025/10}

\definecolor{Gray}{gray}{0.8}

\let\orgautoref\autoref
\renewcommand{\autoref}
{\def\sectionautorefname{Section}%
\def\subsectionautorefname{Section}%
\def\subsubsectionautorefname{Section}%
\orgautoref
}

\usepackage{pifont}
\newcommand{\cmark}{\ding{51}}%
\newcommand{\xmark}{\ding{56}}%

\usepackage{xspace}
\newcommand{\etal}{\textit{et al.}~}
\newcommand{\eg}{\textit{e.g.,}~}
\newcommand{\ie}{\textit{i.e.,}~}

\newcommand{\cf}{\textit{cf.,}~}
\newcommand{\one}{({\em i})\xspace}
\newcommand{\two}{({\em ii})\xspace}
\newcommand{\three}{({\em iii})\xspace}
\newcommand{\four}{({\em iv})\xspace}
\newcommand{\five}{({\em v})\xspace}

\newcommand*\circledScalableFill[4]{\tikz[baseline=(char.base)]{
		\node[shape=circle,draw,inner sep=#1,fill=#4] (char) {\sf #2#3};}}
	
\makeatletter
\renewcommand{\paragraph}[1]{\vspace*{0.03in}\noindent{\bf #1.}\hspace{0.25ex \@plus1ex \@minus.2ex}}
\newcommand{\paragraphNoDot}[1]{\vspace*{0.03in}\noindent{\bf #1}\hspace{0.25ex \@plus1ex \@minus.2ex}}
\makeatother

\begin{document}

\date{}

\title[
On the Potentials of Misusing Transparent DNS Forwarders in Reflective Amplification Attacks]{Forward to Hell?
On the Potentials of Misusing \\Transparent DNS Forwarders in Reflective Amplification Attacks}

\author{Maynard Koch}
 \orcid{0009-0009-3698-1342}
 \affiliation{%
   \institution{TU Dresden}
   \city{Dresden}
   \country{Germany}
 }
 \email{maynard.koch@tu-dresden.de}

\author{Florian Dolzmann}
 \orcid{0009-0002-2591-7264}
 \affiliation{%
   \institution{TU Dresden}
   \city{Dresden}
   \country{Germany}
 }
 \email{florian.dolzmann@mailbox.tu-dresden.de}

 \author{Thomas C. Schmidt}
 \orcid{0000-0002-0956-7885}
 \affiliation{%
   \institution{HAW Hamburg}
   \city{Hamburg}
   \country{Germany}}
 \email{t.schmidt@haw-hamburg.de}

 \author{Matthias W\"ahlisch}
 \orcid{0000-0002-3825-2807}
 \affiliation{%
   \institution{TU Dresden}
   \city{Dresden}
   \country{Germany}
 }
 \email{m.waehlisch@tu-dresden.de}

\definecolor{boxgray}{rgb}{0.93,0.93,0.93}
 \textblockcolor{boxgray}
 \setlength{\TPboxrulesize}{0.7pt}
 \setlength{\TPHorizModule}{\paperwidth}
 \setlength{\TPVertModule}{\paperheight}
 \TPMargin{5pt}
 \begin{textblock}{0.8}(0.1,0.04)
   \noindent
   \footnotesize
   If you refer to this paper, please cite the peer-reviewed publication:  
   Maynard Koch, Florian Dolzmann, Thomas C. Schmidt, and Matthias Wählisch. 2025.
   Forward to Hell? On the Potentials of Misusing Transparent DNS Forwarders in Reflective Amplification Attacks.
   \emph{Proceedings of the 2025 ACM SIGSAC Conference on Computer and Communications Security (CCS '25), October 13--17, 2025, Taipei,Taiwan}. https://doi.org/10.1145/3719027.3765096
\end{textblock}

\begin{abstract}
The DNS infrastructure is infamous for facilitating reflective amplification attacks. Various countermeasures such as server shielding, access control, rate limiting, and protocol restrictions have been implemented. Still, the threat remains throughout the deployment of DNS servers. In this paper, we report on and evaluate the often unnoticed threat that derives from transparent DNS forwarders, a widely deployed, incompletely functional set of DNS components. Transparent DNS forwarders transfer DNS requests without rebuilding packets with correct source addresses. As such, transparent forwarders feed DNS requests into (mainly powerful and anycasted) open recursive resolvers, which thereby can be misused to participate unwillingly in distributed reflective amplification attacks. We show how transparent forwarders raise severe threats to the Internet infrastructure. They easily circumvent rate limiting and achieve an additional, scalable impact via the DNS anycast infrastructure. We empirically verify this scaling behavior up to a factor of 14. Transparent forwarders can also assist in bypassing firewall rules that protect recursive resolvers, making these shielded infrastructure entities part of the global DNS attack surface.  
\end{abstract}


\maketitle

\section{Introduction}
\label{sec:introduction}
\begin{figure}
	\centering
	\includegraphics[width=\linewidth]{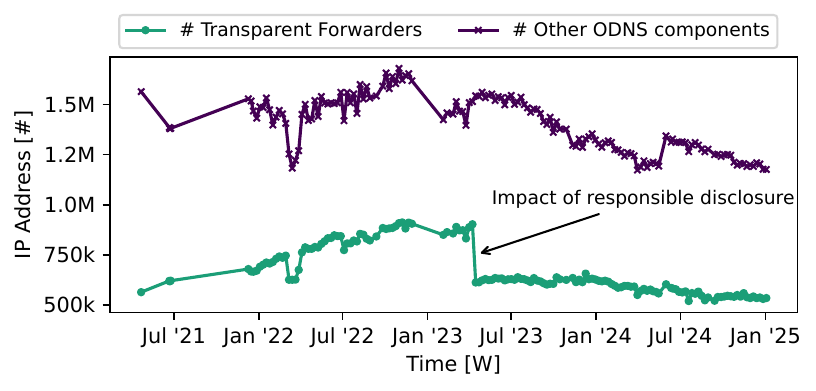}
	\caption{Number of transparent DNS forwarders and other open DNS components per week from June 2021 to January 2025. By sharing our results with network operators, we were able to decrease the impact of transparent forwarders by 30\%.}
	\label{fig:no_odns_components}
\end{figure}

The open DNS (ODNS) infrastructure is a valuable target for attackers to misuse them in DNS reflective amplification attacks. While DNS service providers implement various countermeasures, the major part of the ODNS is unintentionally open to anyone, hence lacking of proper security configurations. 
Originally observed in 2013~\cite{mauch2013nanog} and measured in detail in 2021~\cite{nawrocki2021transparent}, transparent DNS forwarders, which forward DNS requests to their configured resolver without rewriting the source IP address, currently account for 30\% of the ODNS infrastructure, see \autoref{fig:no_odns_components}.
Despite their fundamental attack surface, transparent forwarders still receive little attention.
Several scanning campaigns such as Censys~\cite{censys2017search} and Shadowserver~\cite{shadowserverDNS} do not discover transparent forwarders and thus underestimate the threat potential of the ODNS landscape.
\begin{figure*}[t]
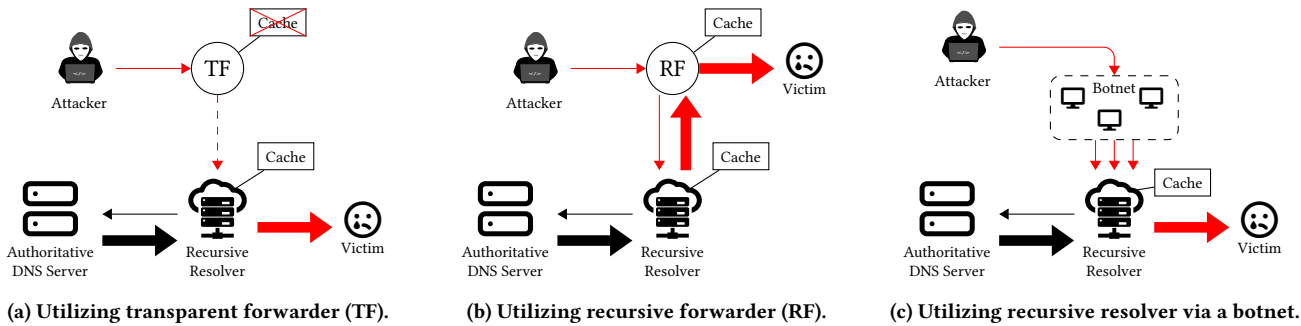

	\centering
	\begin{subfigure}{0.33\linewidth}
		\centering
		\input{figures/odns-infrastructure/tfwd-attack}
		\caption{Utilizing transparent forwarder (TF).}
	\end{subfigure}
\hfill
	\begin{subfigure}{0.33\linewidth}
		\centering
		\input{figures/odns-infrastructure/rfwd-attack}
		\caption{Utilizing recursive forwarder (RF).}
	\end{subfigure}
\hfill
	\begin{subfigure}{0.33\linewidth}
		\centering
		\input{figures/odns-infrastructure/resolver-attack}
		\caption{Utilizing recursive resolver via a botnet.}
	\end{subfigure}
\caption{Data paths in reflective amplification attacks via different open DNS components. Transparent forwarders can scale up attacks as they do not carry amplified traffic and enable distribution without a botnet.}
\label{fig:overview-attack-paths-odns}
\end{figure*}

In this paper, we show that transparent DNS forwarders largely extend the attack surface over recursive resolvers and forwarders. 
Transparent forwarders introduce uniquely open DNS resolvers that send requests using the source IP address of the stub resolver and do not hold states.
Therefore---as new security risks---transparent forwarders enable an attacker to misuse the anycast infrastructure of large recursive resolvers and allow much higher attack rates.
Furthermore, transparent forwarders increase the reachability of recursive resolvers.
We observe 200 shielded resolvers that are exclusively reachable via transparent forwarders.

Transparent forwarders \one scale significantly better during an attack as they do not need to handle the (amplified) DNS responses, \two redirect queries to recursive resolvers that belong to a powerful infrastructure, such as Google or Cloudflare, \three bypass rate limits via orchestration without the need of an extensive botnet, \four~can bypass filters and shields that protect resolver infrastructure, and \five~provide faster response times than  recursive forwarders in countries where they are primarily located.
Our main contributions read:
\begin{enumerate}
	\item We introduce an attacker model, the security risks of transparent forwarders in reflective amplification attacks, and two attack scenarios (\S\ref{sec:attack-model}).

  \item We analyze the vendor and device type landscape of transparent forwarders using banner grabbing and SNMP measurements. MikroTik dominates the current deployments with device classes ranging from constrained CPE devices to powerful core routers (\S\ref{sec:fingerprinting}).

	\item We provide lower bounds of the performance advantages of transparent forwarders over recursive forwarders when executing an attack, and show empirically that transparent forwarders can therefore cause over 14$\times$ more traffic at a potential victim compared to the use of recursive forwarders~(\S\ref{sec:tfwd_vs_rfwd}).

	\item We define a general setup to measure transparent forwarders, rate limits of recursive resolvers, and amplification (\S\ref{sec:experiment-setup}).

	\item We circumvent poorly configured access restrictions of recursive resolvers by utilizing transparent forwarders and analyze their suitability to conduct reflective amplification~attacks~(\S\ref{sec:exploiting-shielded-resolver}).

	\item We orchestrate transparent forwarders to bypass regional rate limits of large DNS anycast providers such as Google, Cloudflare, and OpenDNS (\S\ref{sec:exploiting-anycast-deployments}).
\end{enumerate}

Prior work focuses on revealing transparent forwarders.
To the best of our knowledge, this paper is the first that assesses the threat potential and security risks of transparent forwarders in detail.
We use custom responses instead of custom queries to reduce overhead and impact of our scans on the Internet.
Additionally to our Internet scans, we conduct an end-to-end attack experiment in our lab using a MikroTik router and a DNS server under our control. We use this experiment to illustrate the attack potential and the impact in a real-world scenario.
When presenting our results to network operators, most of them were not aware of transparent forwarders.
For this reason, we believe that a systematic understanding such as that provided in this paper is important.

After summarizing the state of the current ODNS infrastructure and its attack vectors in \autoref{sec:background}, we present our main contributions in Sections \ref{sec:attack-model}--\ref{sec:exploiting-anycast-deployments}.
We discuss our results comprehensively, including providing mitigation options against attacks via transparent forwarders, in \autoref{sec:discussion}.
We conclude in \autoref{sec:conclusion}.
We address ethical concerns in \autoref{sec:ethics} and describe research artifacts of this paper in \autoref{sec:artifacts}.

\section{Background and Related Work}
\label{sec:background}

In this section, we explain technical details of the open DNS infrastructure, describe attack vectors, and discuss related work.

\subsection{Open DNS Infrastructure}
\label{sec:odns-background}

\paragraph{Open DNS (ODNS)}
The open DNS (ODNS) infrastructure comprises all devices that accept DNS requests from any host. 
This open system can be divided into \one \emph{recursive resolvers}, which directly resolve queries by communicating with authoritative DNS servers, and \two \emph{forwarders}, which, instead of resolving, forward queries to a recursive resolver~\cite{schomp2013client,anagnostopoulos2013dns} and therefore introduce dependencies to larger resolver clusters~\cite{wwqll-ocucd-25}. Furthermore, it is important to distinguish between \emph{recursive} and \emph{transparent forwarders}. The main difference is that a transparent forwarder does not rewrite the source IP address of the DNS request, hence no bidirectional DNS transaction between transparent forwarder and recursive resolver takes place, but the packet is forwarded to the configured recursive resolver of the transparent forwarder, causing the recursive resolver to directly respond to the client~\cite{nawrocki2021transparent}.
Over the past decade, the total amount of open DNS devices has decreased from over 25M in 2014~\cite{kuhrer2014exit} down to 1.7M in 2024~\cite{yrs-gmror-24}. 
These devices are often target of attackers misusing them as reflectors for DNS requests with spoofed source IP addresses (\cf \autoref{fig:overview-attack-paths-odns}).

\paragraph{Deployment of transparent DNS forwarders}
\begin{figure*}[ht]
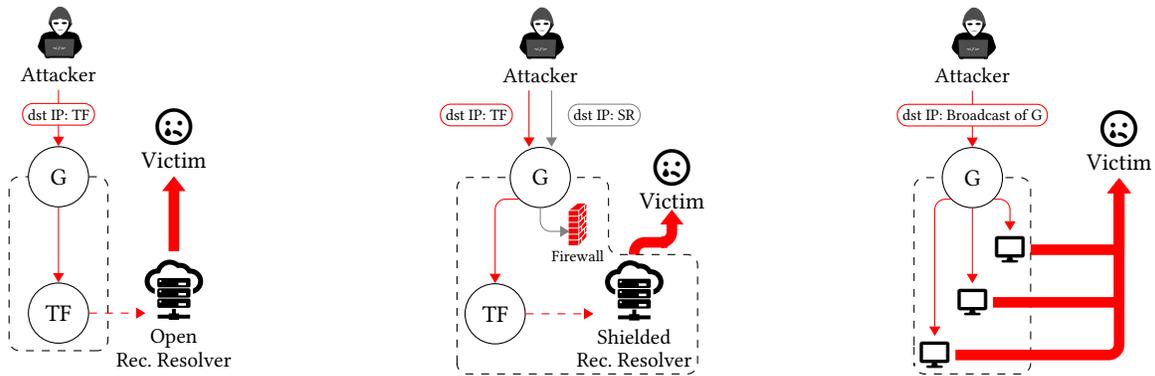

	\centering
	\begin{subfigure}{0.32\textwidth}
		\centering
		\input{figures/tfwd-deployment/tfwd-deployment-open-resolver}
		\caption{An attacker targets a transparent forwarder~(TF) in an unprotected network via the gateway G. The TF relays the query to an open recursive resolver.}
		\label{fig:tfwd-deployment-open-resolver}
	\end{subfigure}
	\hfill
	\begin{subfigure}{0.32\textwidth}
		\centering
		\input{figures/tfwd-deployment/tfwd-deployment-shielded-resolver}
		\caption{An attacker cannot reach the shielded resolver directly (via gateway G), but indirectly using the transparent forwarder.\\}
		\label{fig:tfwd-deployment-shielded-resolver}
	\end{subfigure}
	\hfill
	\begin{subfigure}{0.32\textwidth}
		\centering
		\input{figures/tfwd-deployment/tfwd-deployment-intercepted-query}
		\caption{A misconfigured gateway G takes the role of a TF by forwarding broadcasted DNS requests to the local network. Multiple hosts reply with a DNS response to the victim.}
		\label{fig:tfwd-deployment-intercepted-query}
	\end{subfigure}
	\hfill
	\caption{Deployment scenarios of transparent forwarders in DNS reflection and amplification attacks.}
	\label{fig:tfwd-deployment-scenarios}
\end{figure*}
Originally observed in 2013~\cite{mauch2013nanog}, transparent forwarders received only little attention. A detailed analysis of their deployment was first published in 2021~\cite{nawrocki2021transparent}. These findings show that, while all other ODNS components experienced a major decrease over the past decade, the absolute amount of transparent forwarders remained unchanged.
The definition of a transparent forwarder by Nawrocki \etal~\cite{nawrocki2021transparent} implies three deployment scenarios in which transparent forwarder IP addresses can be misused for DNS reflection and amplification attacks (\cf \autoref{fig:tfwd-deployment-scenarios}).

The most common deployment scenario~\cite{nawrocki2021transparent} is a transparent forwarder \textit{TF} that relays incoming requests to an external open recursive resolver such as Google's 8.8.8.8, which then replies directly to the victim with the amplified response (\cf \autoref{fig:tfwd-deployment-open-resolver}).

Some transparent forwarders allow third parties to expose and misuse DNS recursive resolvers that are meant to be accessible for a predefined range of customer IP~addresses only. We call them \emph{shielded resolvers} as they are shielded from the global network by firewall configuration at AS~borders but are publicly accessible within the AS. When an attacker queries that resolver directly, it does not trigger any response as the request is already filtered at the network borders. However, if the attacker sends its query via the transparent forwarder \textit{TF}, the request is relayed transparently to the shielded resolver \textit{SR}, triggering a valid DNS response (\cf \autoref{fig:tfwd-deployment-shielded-resolver}).

Finally, a few misconfigured routers forward broadcast traffic from remote locations into their network, triggering multiple DNS~responses for a single request, which are then sent to the victim (\cf \autoref{fig:tfwd-deployment-intercepted-query}).
For cases in which the router replies to the request, responses may appear as DNS interceptions~\cite{randall2021homejack,liu2018interception,anonymous2014greatfirewall}, even though the router only replies to broadcast packets following RFC 919~\cite{RFC-919}.
As the initial destination and the replying source IP addresses do not match, they are classified as transparent forwarders.

It is worth mentioning that transparent DNS forwarders are not actively deployed to serve a global range of customers.
They are inadvertently exposed to the public, most likely without the operator being aware of the misconfigured state.
The existence of a transparent forwarder in a network implies the lack of network ingress filtering, which is highly recommended to be deployed to defeat DDoS~\cite{RFC-2827,hnbkh-adeci-24}. 

\subsection{Open DNS Attack Vectors}
\label{sec:odns-attack-vectors}
Two methods are most common to misuse open DNS devices, \one~inflating the response packet size, \eg by querying DNS resource records that generate large responses or poisoning caches with arbitrary records~\cite{zlpyz-potfc-20,man2020poisening}, and \two generating large numbers of responses by triggering chain reactions or using misconfigured resolvers that generate more than a single response per request, or orchestrating a large number of reflectors.

\paragraph{Amplification by packet size}
In DNS reflective amplification attacks, the attacker aims to trigger responses at resolvers that are much larger than the initial request~\cite{rowrs-adads-15}. Amplification factors of 40$\times$ to 60$\times$ are achievable through EDNS0 and DNSSEC signed domains~\cite{anagnostopoulos2013dns,yazdani2022matter,r-ahrnp-14,msk-tbbft-17}. Nawrocki \etal~\cite{njsw-fsdat-21} showed that in cases of DNSSEC key rollovers amplification factors can surpass 80$\times$. In the remainder of this paper, we use an amplification factor of 40$\times$ as a lower bound for the attack potential when misusing ODNS components. The key findings of Nawrocki \etal include that \one~attackers often misuse legitimate \textit{.gov} names as they already provide a high amplification factor because they are DNSSEC signed.
Yazdani \etal analyzed the attack potential of open DNS resolvers in detail, showing that removing the 20\% most potent amplifiers could reduce the attack potential up to 80\%~\cite{yazdani2022matter}.

\paragraph{Amplification by number of responses}
While attackers often misuse domains that trigger large responses in terms of packet size, there is a high abuse potential for reflectors that trigger millions of responses for a single request. These response floods are caused by routing loops and misconfigured middleboxes~\cite{nkd-rlmad-22} and also by \textit{Echoing Resolvers}~\cite{ynhkj-hedrs-23}, which are devices that respond with more than a single reply per request. While the number of such routing loops and misconfigured devices seems to be relatively low, Nawrocki \etal~\cite{njsw-fsdat-21} showed that even 100 affected devices are sufficient for real-world attacks. In most attacks, only 10 to 1000 reflectors are used, emphasizing that the open DNS infrastructure, with (currently) 1.7M~\cite{yrs-gmror-24,nawrocki2021scanning} active IP addresses, is highly vulnerable to such attacks. 
Hundreds of thousands of such resolvers exist in well-provisioned networks~\cite{yhhrd-mspcd-22}, making them a reliable part of DDoS attacks. 
Besides response packet size, IP-address churn plays a fundamental role in the selection of resolvers misused in attacks~\cite{yrs-gmror-24}; open resolvers with a high churn are less likely to be used in an attack. We show that a carefully selected target list of stable transparent forwarders already exceeds the number of resolvers misused in most attacks. The minority of such devices unmistakably indicate their service in hostnames, thus are exposed intentionally to the public~\cite{yjs-srieo-24}. Furthermore, Yazdani \etal show that only 1\% of open resolvers are observed in more than 95\% of the scans within their one-year measurement period, which is still a non-negligible amount of around 20K to 30K devices to be misused as reflectors in DDoS attacks.

\section{Problem Statement}
\label{sec:attack-model}

In this section, we introduce our attacker model to exploit transparent forwarders, explain their advantages when executing amplification attacks, and compare with recursive forwarders.
We illustrate the potentials of transparent forwarders in two concrete scenarios.

\subsection{Attacker Model}
The capabilities required of an attacker are relatively low.
We assume an off-path attacker that is able to send spoofed IP~packets via its upstream provider.
This means an attacker does not need to eavesdrop, intercept, drop, or alter packets along the communication path, but its provider allows for topologically incorrect source addresses.
Although network operators are aiming for deploying ingress filtering since many years~\cite{RFC-2827}, networks that allow for spoofing are quite common~\cite{caidaSpoofer,kuhrer2014exit}.
Most importantly, the attacker in our scenarios does not need to control a botnet to achieve a distributed~attack.

\subsection{Security Risks of Transparent Forwarders in Reflective Amplification Attacks}
Misusing transparent DNS~forwarder leads to a number of advantages when executing reflective amplification attacks compared to public resolvers.

\paragraph{Availability of additional attack infrastructure}
Transparent forwarders give an attacker additional possibilities to exploit the DNS infrastructure by the sheer amount they contribute to the DNS ecosystem. 
About 33\% of all open DNS components are transparent forwarders.

\paragraph{Higher scalability}
Transparent forwarders experience less load because they only need to handle the DNS request, but not the (amplified) response.
Since the response is much larger than the request in an amplification attack, a transparent forwarder requires only a fraction of the bandwidth and computing power to facilitate a similar sized attack compared to a recursive forwarder or resolver.
As such, more devices are available for maximizing the attack volume instead of neutralizing potential resource exhaustion of ODNS components.

While recursive forwarders and recursive resolvers may employ caching, transparent forwarders provide no caching.
The advantage of caching, however, is negligible considering simple name-based rate limiting.

\paragraph{More powerful infrastructure}
Transparent forwarders redirect the resource intensive recursive workload of DNS resolution to recursive resolvers that belong to a powerful infrastructure.
Recent measurements show that a recursive resolver belonging to either Google or Cloudflare is configured on 76\% of all transparent forwarders.
An attacker that simply bases its attacks on recursive resolvers in general may likely target less powerful resolvers (\eg CPE~devices).

\paragraph{Bypass rate limiting}
Transparent forwarders make it easier to bypass rate limits without deploying a botnet.
A botnet would be required in case of public recursive resolvers or recursive forwarders to send queries from multiple IP~addresses but requires compromising third party hosts---a relatively strong assumption.
Instead, instrumenting transparent forwarders requires only a single host with IP~spoofing capabilities.

\paragraph{Faster response times}
\begin{figure}[t]
     \begin{center}
     \includegraphics[width=1\columnwidth]{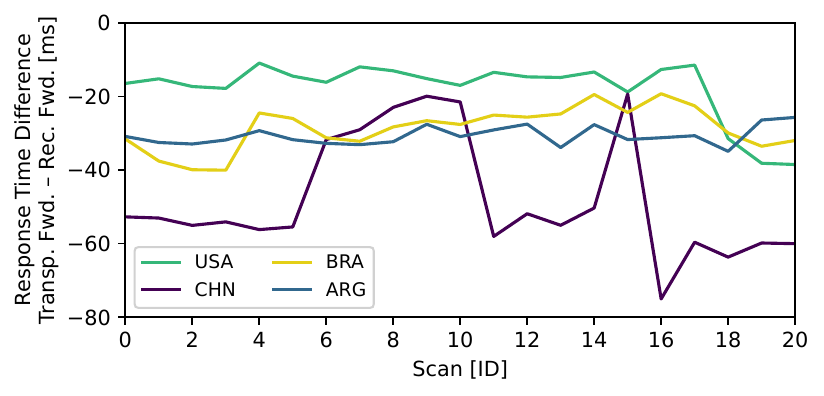}
     \caption{Differences of median response times per scan between transparent and recursive forwarders for countries with a high number of transparent forwarders. Transparent forwarders reply faster.}
     \label{fig:response_times}
     \end{center}
\end{figure}
Additionally, transparent forwarders show faster response times in the majority of countries where they are primarily located.
\autoref{fig:response_times} shows the difference of the median response time between transparent and recursive forwarder for four countries with a high amount of transparent forwarders deployed over the course of 5 months.
Even though the response times seem to fluctuate significantly, a consistent trend is visible, with faster responses in Brazil, Argentina, China and the USA, with up to 60ms and on average around 20-40ms faster replies compared to recursive forwarders.
The requested resource record is hereby always uncached. 
By additionally comparing the response times of one unchanged and one possibly cached request, depending on recursive forwarder configuration, we observe around 25\% of all recursive forwarders with at least 20ms faster replies, while the other 75\% show no significant difference.
We therefore suppose that response times of recursive forwarders are much more subject to routing changes, network congestion and load of the forwarder than cache hits and misses.
In choosing transparent over recursive forwarders that respond faster and might therefore be under less load an amplification attack can be performed more efficiently.

\subsection{Scenario 1: Exploiting Shielded Resolvers}
In this scenario, an attacker exploits shielded resolvers (see \autoref{sec:background}). \ie recursive resolvers that are protected by a firewall but the firewall can be bypassed by a transparent forwarder. 
The scenario is visualized in \autoref{fig:tfwd-deployment-shielded-resolver}. 
Since shielded resolvers are protected, we expect their system configurations to be less restrictive in terms of response rate limiting and query types.
Based on our measurements, successful queries of \texttt{ANY}-type, \texttt{TXT}-type, and DNSSEC-enabled records are possible.
Combining both restricted rate limiting and flexible query types allows for fast and highly amplified replies towards a victim.

Transparent forwarders are not the only open DNS components that can bypass firewalls.
Shielded resolvers can also be accessed indirectly through recursive forwarders.
In contrast to recursive forwarders, however, transparent forwarders are less likely to be affected by rate limiting of DNS~queries because transparent forwarders aim to minimize states and assume any query to be legitimate.

Deploying rate limiting on the recursive resolver in such shielded setups is neither necessary nor helpful for two reasons.
A rate limiting algorithm filters based on the source IP~address of the requesting client.
If a recursive resolver receives queries from a recursive forwarder triggered by different stub resolvers, the source IP~address would be the same for subsequent queries.
Applying rate limiting on the (firewall-protected) recursive resolvers would lead to many false negatives, which is the reason why we argue that rate limiting will be deployed on recursive forwarders in such scenarios.
Furthermore, considering that the recursive resolver would normally be protected by the firewall and only indirectly accessible over the already filtered recursive forwarder, there might be little incentives to enable filtering on such recursive~resolvers.

\subsection{Scenario 2: Exploiting Anycast\\ Infrastructures}
\label{sec:scenario-anycast}
\begin{table}%
	\centering
	\caption{Top 10 public DNS resolvers used by transparent forwarders. Most transparent forwarders relay queries to Google (67\%) or Cloudflare (9\%).}
	\label{tab:top_10_resolvers}
	\begin{tabular}{llrr}
		\toprule
		\multicolumn{2}{c}{Public Resolver} & \multicolumn{2}{c}{{\makecell[b]{Transparent Forwarders \\ Using Public Resolvers}}} \\
		\cmidrule(r){1-2}
		\cmidrule{3-4}
		IP Address & Provider & [\#] & [\%] \\
		\midrule
    8.8.8.8 & Google & \num{341447} & 64.25 \\
  1.1.1.1 & Cloudflare & \num{48313} & 9.09 \\
208.67.222.222 & OpenDNS & \num{14464} & 2.72 \\
8.8.4.4 & Google & \num{14115} & 2.66 \\
223.29.207.110 & Meghbela & \num{11789} & 2.22 \\
83.220.169.155 & Comss.one DNS & 2047 & 0.39 \\
178.233.140.109 & Turksat & 1790 & 0.34 \\
203.147.91.2 & Meghbela & 1634 & 0.31 \\
1.0.0.1 & Cloudflare & 1196 & 0.23 \\
103.88.88.88 & DNS Bersama & 1007 & 0.19 \\	
		\bottomrule
	\end{tabular}
\end{table}
In this scenario, we focus on taking advantage of the anycast deployments of popular public DNS resolver infrastructures operated by large-scale providers, \eg Google and Cloudflare.
The ODNS services of those providers are used by the majority of transparent forwarders, see \autoref{tab:top_10_resolvers}. 

The anycast infrastructures of a large-scale provider brings the advantage of a well-maintained, highly distributed backend that easily scales with an increasing number of requests.
Being able to exploit such infrastructures for reflective amplification attacks poses a major threat.

\begin{figure}%
	\input{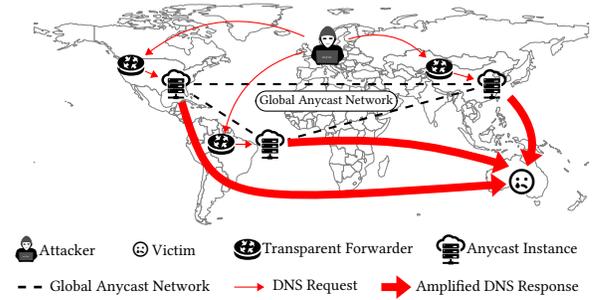}
	\caption{Using transparent forwarders to launch a globally distributed DNS amplification attack via large public DNS providers with anycast deployment such as Google.}
	\label{fig:scenario_anycast}
\end{figure}
The idea is to use DNS forwarders at different locations to reach globally spread anycast instances of a recursive DNS provider such as Google, as shown in \autoref{fig:scenario_anycast}.
Our goal is to explore whether response rate limits, which we locally see on a single resolver instance, can be surpassed by concurrently sending globally distributed DNS requests through forwarders to the different recursive resolver endpoints of the anycast network.
As a baseline, we therefore look at the rate limiting behavior of popular resolvers at three distinct points of presence by requesting the resolvers directly without the use of any forwarders. We compare this data to the results of using multiple transparent forwarders simultaneously. Then, we estimate the threat potential based on the size of the anycast infrastructure accessible via forwarders.

\section{Fingerprinting Transparent Forwarders}
\label{sec:fingerprinting}
In this section, we present our results on fingerprinting and classifying transparent forwarders.

\paragraph{Method}
The starting point of our measurements are all transparent forwarders that we identified on the Internet (see \autoref{sec:principle-setup} for details on the measurement method).
We use ZGrab~\cite{dambh-sebiw-15} and a fast SNMP scanner~\cite{onesixtyone} to retrieve systems information about these transparent forwarders.
To extract information such as operating system~(OS), device model, and device vendor, we apply regular expressions.
Finally, we group devices for which we successfully gathered information to understand structural properties.

Using ZGrab, we scan for common Web ports (TCP/80, TCP/8080, TCP/443) and SSH (TCP/22). 
Some webpages, however, are delivered slowly due to geographic distances or constrained hardware resources of the servers, which conflict with timeouts of ZGrab. 
Therefore, we use Selenium, a browser automation tool, to scan all devices that appear to be open but send no HTTP~header information in the response. We set a timeout of 20s to retrieve HTML~data. 

\begin{table}
    \centering
    \caption{Fingerprinting results by device type and vendor. The majority of transparent forwarders are routers, but some are also identified as network video recorders. Most of the MikroTik routers are powerful core devices.}\label{tab:fingerprint-results}
    \begin{tabular}{llr}
        \toprule
        Device Type & Vendor & Devices [\#]\\
        \midrule
        Router  & MikroTik (Core) & 5569 \\
                & MikroTik (CPE)  & 4362 \\
                & TP-Link  & 728  \\
                & Ubiquiti & 663  \\
                & Fortinet & 252  \\
                & ZTE      & 200  \\
                & Cisco    & 104  \\
                & Zyxel    & 102  \\
                & Huawei   & 58   \\
                & D-Link   & 24   \\
                \cmidrule{2-3}
                & Other    & 114  \\
        \midrule
        Network Video Recorder & HikVision & 871\\
            & UNV       & 25   \\
        \bottomrule
    \end{tabular}
\end{table}

\paragraph{Results}
We are able to fingerprint \num{13072} (2.5\%) of $\approx$\num{530000} transparent forwarders.
Even though this number is much lower than the overall number of transparent forwarders, global applicability still holds since we learn more details about this global subset, \eg to derive performance properties of the potential attack infrastructure.

The majority of the devices that we can fingerprint are MikroTik routers (76\%).
Those MikroTik devices can be divided into core routers, which are powerful devices such as \textit{CCR2116-12G-4S+} or \textit{CCR1036-8G-2S+}, and CPE devices such as \textit{RB750Gr3} or \textit{RB760iGS}.
We observe routers as the major type of transparent forwarders, however, we also discover network video recorders such as HikVision or UNV IP-cameras.
While we are only able to map 2.5\% of the transparent forwarder landscape to a vendor and device type, it is clearly visible that transparent DNS forwarder behavior is not limited to MikroTik. Although the fingerprinting is limited to a small sample, the identified devices are distributed in 1544 ASNs over 103 countries, therefore indicating a global trend.
Furthermore, transparent forwarders cover a broad range of devices, from constrained CPE up to powerful core~routers.
We summarize our results in \autoref{tab:fingerprint-results}.

\section{Comparing Performance of Transparent Forwarders and Recursive Forwarders}
\label{sec:tfwd_vs_rfwd}

Both transparent forwarders and recursive forwarders allow to bypass firewalls that protect shielded resolvers.
Transparent forwarders, however, pose the unique security risk that they do not need to handle the amplified reply, which increases scalability of the threat landscape (see \autoref{sec:attack-model}).
Comparing the limits of both types of DNS forwarders based on an Internet-wide measurement study would conflict with ethical concerns and is therefore not in scope of this work.
Instead, we provide a numerical analysis bolstered by empirical data gathered in a lab experiment to reflect structural properties.

\subsection{Method}
The goal is to determine how much more efficiently a transparent forwarder can facilitate an attack compared to a recursive forwarder.

\paragraph{Numerical analysis}
In the following calculations, we assume a 70 byte large DNS query packet and an amplification factor of at least 40$\times$ (see \autoref{sec:odns-attack-vectors}).
This conservative configuration leads to a response packet size of 2800 byte.
The response packet, hence, needs to be fragmented into two~packets, considering the common MTU of 1500 byte.
This introduces additional processing load and delay because fragmented packets must be reassembled before they can be processed on the application layer.

We base our analysis on MikroTik routers, a major vendor of devices deployed in the wild running transparent forwarders~(see \autoref{sec:fingerprinting}).
We consider two router models representing two different performance classes, a less and a more powerful device, based on the most common devices found by our fingerprinting.
The MikroTik~RB750Gr3 is a five~port Gigabit MIPS device commonly used in small businesses or home installations.
The MikroTik~CCR1036-8G-2S+ is a discontinued eight~port Gigabit device featuring a 36-core CPU, usually serving larger corporate networks.

We assume the constrained router to be connected asymmetrically with 100Mbit/s downstream and 50Mbit/s upstream, which was the global median in February 2025 for fixed broadband as determined by Ookla~\cite{speedteststats}, and the powerful router with a symmetrical connection of 1 Gigabit/s to the Internet.
The basic performance characteristics in terms of possible packet rates are provided by the vendor~\cite{mikrotikrb750gr3,mikrotikccr10368g2s}.
In our calculation, we apply the slow path packet rate when a router processes a packet since any DNS packet needs to be processed by the CPU.
Given a common deployment setup (\eg basic access control lists), the powerful router can process up to 3050k small sized (64 Byte) and 1825k large (1518 Byte) packets per second and the less powerful router $\approx$64k~packets per second independent of packet size.
Depending on the router model the processing time seems to be also limited by packet size and not just the compute power.
These performance characteristics indicate that an attacker might need to adjust requests to achieve a maximum attack rate, but this comes at the trade-off of reducing the amplification factor and is most likely counterproductive, as can also be seen in the empirical data provided later on.

We can now determine the maximum rate of DNS queries the router may be able to forward and therefore also the maximum attack traffic a victim would receive. 
This calculation provides an upper limit since we make some assumptions for the sake of simplicity.
First, we ignore side traffic load on the router in our model.
Such load depends on specific setups and is out of scope.
Furthermore, we assume the same processing time for DNS and other application layer protocols.
We also conjecture that a recursive forwarder will spend more time in slow path than a transparent forwarder due to additional DNS query logging, filter list lookups, and cache operations. We do not expect such actions for a transparent forwarder because DNS packets pass transparent forwarders unchanged, except for a change of the destination IP~address and the lookup of the recursive resolver.
Overall, our assumptions ensure a fair comparison, slightly overestimating the resources available on recursive forwarders.

\paragraph{Testbed} 
To empirically verify our observations, we measure the amount of traffic that can be generated based on a testbed using a MikroTik RB750Gr3 router, an authoritative nameserver, and a DNS client.
The authoritative nameserver is powered by an Intel i5 6500, running an instance of Bind 9 to serve predefined DNS payloads with different sizes.
We conduct two experiments, \one the router is configured as a regular recursive DNS forwarder and \two as transparent forwarder.

\subsection{Results}
\autoref{tab:router_comp} summarizes the results of our numerical evaluation based on two concrete router implementations.

\begin{table}%
	\setlength{\tabcolsep}{2pt}
	\centering
	\caption{Performance comparison of recursive and transparent forwarders running on two different router models.}
	\label{tab:router_comp}
	\begin{tabular}{lrr}
		\toprule
		& \multicolumn{2}{c}{MikroTik Router Model} \\
		\cmidrule{2-3}
		& RB750Gr3 & CCR1036-8G-2S+ \\
		& (constrained) & (powerful) \\
		\midrule

		\textbf{Configuration}
		\\
		\quad \textbf{Uplink} & 50 Mbit/s & 1 Gbit/s \\
		\quad \textbf{Downlink} & 100 Mbit/s & 1 Gbit/s \\
        \quad \textbf{Max. packet rate} & $\sim$64k pkts/s& 1825k-3050k pkts/s\\
		\quad \textbf{Effective query rate} &&\\
        \qquad Transparent forwarder & 64k queries/s & 1785k queries/s  \\
        \qquad Recursive forwarder & 2.2k queries/s &  44.6k queries/s\\
		\midrule
		\multicolumn{3}{l}{\textbf{Results}} %
		\\
        \multicolumn{3}{l}{\quad \textbf{Transparent forwarder}}\\
		\qquad Traffic router $\rightarrow$ Internet & 35.8 Mbit/s & 1 Gbit/s \\
		\qquad Traffic Internet $\rightarrow$ victim & 1.43 Gbit/s & 40 Gbit/s \\
        \multicolumn{3}{l}{\quad \textbf{Recursive forwarder}} \\
		\qquad Traffic Internet $\rightarrow$ router & 1.25 Mbit/s &  25 Mbit/s \\
		\qquad Traffic router $\rightarrow$ victim & 50 Mbit/s &  1 Gbit/s \\
        \quad \textbf{Ratio transp./rec.} & $\sim$29 & $\sim$40 \\
		\bottomrule
	\end{tabular}
\end{table}

\paragraph{Constrained router}
A transparent forwarder only needs to handle the DNS request and not the response, and therefore processes only one packet of a full DNS exchange. The upper limit of queries per second (qps) is equal to the maximum packet rate. In our setup, the data rate triggered by an attacker via a transparent forwarder can be estimated to be 35.8~Mbit/s, which is well below the connection limit of 50~Mbit/s, clearly not overloading the misused infrastructure.
With the conservative amplification factor of 40$\times$, an attacker can achieve up to 1.43~Gbit/s traffic using a single transparent forwarder, though, since the attack traffic is directly sent from the (more powerful) recursive resolvers to the victim.

This picture changes when misusing a recursive forwarder to conduct an attack.
The effective query rate via a recursive forwarder will only be a third of the maximum packet rate since a recursive forwarder handles the request (1~packet) and response (2~packets), resulting in theoretical attack traffic of 480~Mbit/s sent from the router, which is well above the maximum uplink speed of 50~Mbit/s. We see that the router in transparent forwarder configuration could be used to create roughly up to 29$\times$ the attack traffic of a recursive forwarder.

\paragraph{Powerful router}
The more powerful MikroTik CCR1036-8G-2S+ can achieve a maximum packet rate of between 1825k and 3050k packets per second in slow path mode, depending on packet size. At this packet rate, the router is well above the limit of the uplink of 1Gbit/s when operating as a recursive forwarder. The maximum traffic at the victim end an attacker could produce is 1Gbit/s, the connection speed.
Contrary to this, an attacker misusing a transparent forwarder of the same hardware class can abuse the full Gigabit in bandwidth while the victim receives 40~Gbit/s of traffic, since the resources of the more powerful router are not yet exhausted.

\begin{figure}
	\centering
	\includegraphics[width=\linewidth]{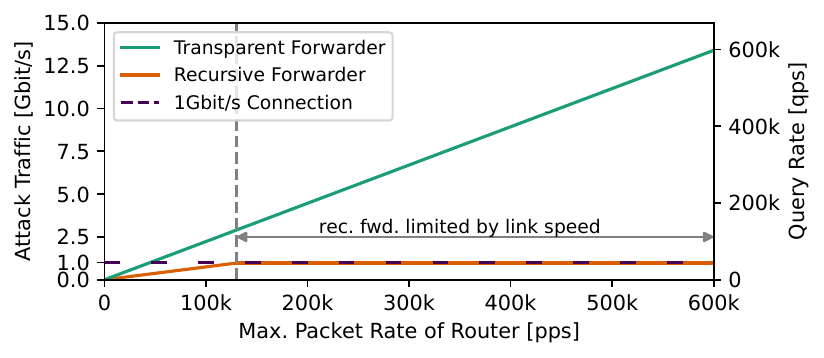}
	\caption{Amount of attack traffic that can be achieved using a transparent forwarder compared to a recursive forwarder, depending on the packet rate a router provides. We assume an uplink of 1~Gbit/s and an amplification factor of 40$\times$.}
	\label{fig:tfwd-vs-rfwd-victim-traffic}
\end{figure}

\paragraph{Generalization}
A fundamental difference between a transparent forwarder and recursive forwarder is that a transparent forwarder is not involved in handling the amplified traffic.
This has direct impact on the scalability when misusing one of the two ODNS~components. 
Even if a recursive forwarder exhibits higher processing capabilities than previously assumed in our assessment, a recursive forwarder is quickly limited by its uplink capacity.
\autoref{fig:tfwd-vs-rfwd-victim-traffic} shows the principal amount of attack volume that can be achieved, independently of specific hardware.
It is clearly visible that the attack volume via a recursive forwarder saturates when the available bandwidth is exhausted.
Before reaching this saturation point, the limiting factor is essentially the compute power of the router, and the transparent forwarder therefore is able to create roughly 3$\times$ more attack traffic than a recursive forwarder.
After the point of saturation this ratio could keep growing to a maximum of roughly the DNS amplification factor, if not limited otherwise by the recursive resolver or its connection~speed.

\begin{figure}
	\centering
	\includegraphics[width=\linewidth]{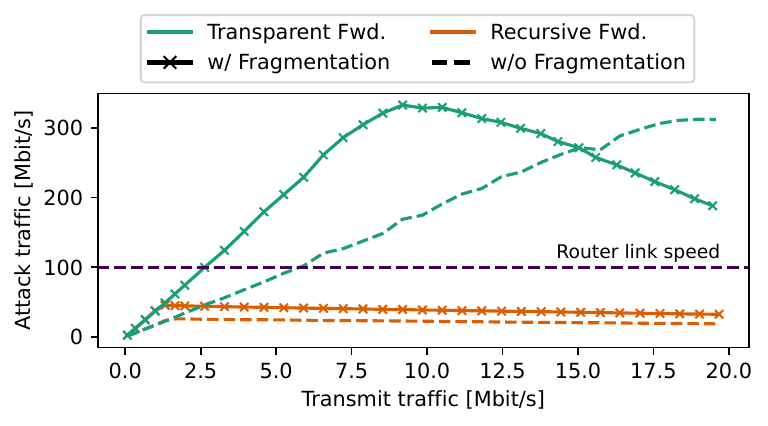}
  \caption{Empirical results using the MikroTik RB750Gr3 router in our testbed. We limited the router link speed to 100MBit/s to match with our assumptions made in~\autoref{tab:router_comp}.}
	\label{fig:mikrotik-comparison-testbed}
\end{figure}

\paragraph{Testbed results}
We use our testbed to practically verify our observations. While we previously assumed that a recursive forwarder would be limited by its link speed, our testbed shows that the \textit{RB750Gr3} router already runs into resource limitations at 1.5Mbit/s of query traffic, resulting in $\approx$50MBit/s attack traffic at the victim, see \autoref{fig:mikrotik-comparison-testbed}. In contrast, when configured as a transparent forwarder, we reach up to 320MBit/s at the victim without running into bandwidth limitations on the transparent forwarder side, highlighting the advantage of transparent forwarders over recursive forwarders in DDoS amplification attacks.

\section{Method and Setup of our Internet-wide Measurements}
\label{sec:experiment-setup}

In this section, we explain our measurement setup to analyze the behavior of transparent forwarders in the wild.
Our experiments are designed to shed light on an attacker that aims to exploit both shielded resolvers and the anycast infrastructure of common public recursive resolvers used by transparent forwarders.
We present the results of these experiments in
\autoref{sec:exploiting-shielded-resolver} and
\autoref{sec:exploiting-anycast-deployments}, and used this data as input source for our analysis in \autoref{fig:no_odns_components}.

Our experiment design is guided by the observation that an attacker does not only try to increase the amplification factor of a single DNS~request but also the amount of data per time towards the victim (\ie data rate). %
Therefore, we do not only implement experiments to exploit shielded resolvers and the anycast deployment of the backend infrastructure in principle but also conduct measurements that reveal specific rate limiting deployments.
It is worth noting that we carefully conduct our experiments such that we do not harm any infrastructure.
This consideration limits our ability to fully explore the attack potential introduced by transparent forwarders since we do not reach the actual amount of attack volume a malicious user could achieve.
We discuss further ethics considerations in detail in \autoref{sec:ethics}.

\subsection{Principle Measurement Method and Setup}
\label{sec:principle-setup}

We measure the open DNS infrastructure since 2021 by weekly scanning the global IPv4 address space, implementing the method described in~\cite{nawrocki2021transparent}.
Our scanner is located in Europe and connected to a university network, unaffected by any university firewall.
It uses a custom \textit{GO}~implementation to scan at scale.
Unless otherwise noted, we will focus on measurements from January~1~to~December~31,~2024.

To identify transparent forwarders, we run an authoritative DNS nameserver operating a zone under our control.
Our server returns two custom \texttt{A} records, \one~a static control record to detect manipulations and \two the IP address of the device that connected to our nameserver ($A_{resolver}$).
We compare the probed IP address~($IP_{target}$) with the replying IP address~($IP_{response}$) combined with a unique triple of (\textit{src port, dst port, dns id}).
Therefore, we can match requests with replies that originate from a different IP address.
We classify open DNS components as follows:
\begin{description}
  \item[Transparent Forwarder] if 
  \newline $ IP_{response} \neq A_{resolver} \, \land \, IP_{target} \neq IP_{response} $

  \item[Recursive Forwarder] if
  \newline $ IP_{response} \neq A_{resolver} \, \land \, IP_{target} = IP_{response} $

  \item[Recursive Resolver] if
  \newline $ IP_{response} = A_{resolver} \, \land \, IP_{target} = IP_{response} $
\end{description}

\begin{figure*}%
	\input{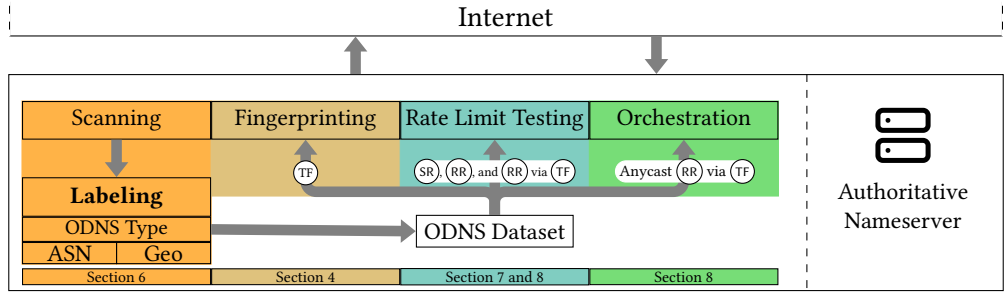}
  \caption{Overview of our measurement infrastructure, scanning setup, and processing. We operate a scanning node as well as an authoritative nameserver to conduct our controlled, active experiments in the wild. The section labels at the bottom indicate which section addresses the highlighted aspect. We fingerprint transparent forwarders (TF) in \autoref{sec:fingerprinting}. We test the rate limit of shielded resolvers (SR) via TF in \autoref{sec:exploiting-shielded-resolver} and the rate limit of open recursive resolvers (RR) \one directly and \two via TF in \autoref{sec:exploiting-anycast-deployments}. We orchestrate anycast resolvers using the infrastructure provided by transparent forwarders in \autoref{sec:exploiting-anycast-deployments}.}
	\label{fig:overview-scanning-setup}
\end{figure*}

The result of each ODNS measurement is a set of IP~addresses that includes a label for each IP~address describing the ODNS~type.
Additionally, we add labels referring to the location and autonomous system of each IP~address.
We use the free MaxMind GeoIP database~\cite{maxmind-geolitecountry} to geolocate scan sources and the freely available RouteViews dataset~\cite{routeviews} to map IPv4 addresses to autonomous~systems.
\autoref{fig:overview-scanning-setup} visualizes our setup.

\subsection{Further Details}
\label{sec:measurement-setup-details}

The principle method and setup described in \autoref{sec:principle-setup} creates the input dataset (\ie IP~addresses of transparent forwarders, recursive forwarders, and recursive resolvers) for our measurements to exploit shielded resolvers as well as the anycast infrastructure by misusing transparent forwarders.
We describe the methods behind those measurements~now.

\paragraph{Measuring rate limits}
To measure DNS response rate limits of recursive resolvers, we start by requesting DNS \texttt{A} records at a slow packet rate of 50~packets per second, iteratively increasing the sending rate up to a maximum of 3000~packets per second in a staircase-like fashion.
We stop our tests at a maximum of 3000~packets per second because we deem this sufficient to understand whether the targeted resolver has a rate limit in place or not. We regard anything above this maximum rate as potentially unlimited, as there seems to be little reason to configure a rate limit this high.

We use a staircase pattern to increase the sending rate because this allows us to detect potential rate limits early during our measurements.
If a rate limit appears in place, we stop the test immediately at that point, so we will not continue sending unnecessary requests.
Additionally, we limit the maximum duration per tested rate to two seconds.

To ensure consistent analysis, we need to cope with churn of IP~addresses and reliably identify replies triggered by our measurements.
Therefore, we test multiple targets (\ie IP~addresses of open transparent forwarders) at the same time and carry meta data in a common header field. %
This meta data is represented by a unique UDP source port assigned to each target.
Any reply by a recursive forwarder will include this port as the destination port, allowing to map the reply to the original forwarder IP~address.

Regarding the names that we request, we use the same name for multiple requests, as well as changing names to analyze rate limiting that is based on names or source IP~addresses.
In both cases, the name is under our control and we have access to the authoritative name servers, \eg to verify log files. 

It is important to note that, in this setup, a drop or limit in response rate does not necessarily indicate that the resolver used by a transparent forwarder is inherently limited or applying a response rate limit. There are a several reasons why this may not always be the case. 
First, we cannot be sure that all of the requests we send at high packet rates are actually passed on by a transparent forwarder to its upstream resolver or whether some of the requests might already have been lost. 
Second, even if querying \texttt{A} records at somewhat higher packet rates is permitted, limits might be in place for queries that trigger larger sized payloads, \eg due to DNSSEC, \texttt{ANY}, or \texttt{TXT} queries. We do not test rate limits of record types other than \texttt{A}, as we deem this unnecessary and potentially damaging.

\paragraph{Measuring DNS amplification}
To determine support for amplifying queries, we probe for DNS \texttt{ANY} and DNSSEC support. We choose a popular domain name, which is most likely to be contained in a resolver cache, and therefore provide a large response, such as \textit{google.com}. We send one DNS \texttt{ANY} request per resolver address and await its response. 
Similarly, to test for DNSSEC, we select a domain name as query name that is very likely secured by DNSSEC.
Domains under \textit{.gov} are good candidates, see \autoref{sec:odns-attack-vectors}.

\paragraph{Measuring shielded resolvers}
When we conduct our regular scans, we find two types of recursive resolvers.
\one~A recursive resolver that sends a DNS request to our authoritative DNS server when we target the IP~address of this resolver.
\two~A recursive resolver that sends a DNS request to our authoritative DNS server when we target the IP~address of a transparent forwarder.
We label a recursive resolver as shielded resolver if this resolver is \emph{only} visible in the second case, because this resolver is only reachable via a transparent forwarder and shielded otherwise.

\section{Exploiting Shielded Resolvers}
\label{sec:exploiting-shielded-resolver}
\begin{figure}
	\centering
	\includegraphics[width=\linewidth]{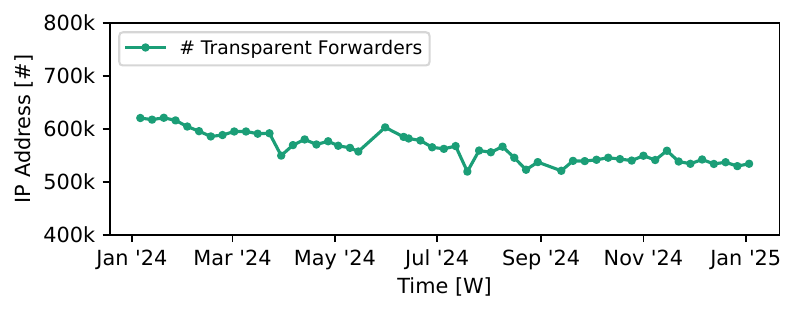}
	\caption{Number of Transparent Forwarders in 2024.}
	\label{fig:no_tfwd}
\end{figure}

\begin{table}
	\centering
    \caption{Top 5 autonomous systems hosting shielded resolvers and the number of transparent forwarders using these resolvers on January~6,~2024.}
	\begin{tabular}{lrcr}
		\toprule
		\multicolumn{2}{c}{Shielded Resolvers} & $\leftarrow$ & \makecell[r]{Transparent Forwarders}
		\\
		\cmidrule(rl){1-2}
		\cmidrule(rl){4-4}
		AS Number (Name)  & \makecell[r]{{[\#]}} & & \makecell[r]{{[\#]}}
        \\
		\midrule
    4812 (China Telecom) & 9880 & & \num{17923} 
    \\
    5769 (Videotron) & 4256 & & 8377 
    \\
    209 (Lumen) & 2446 & & 5576 
    \\
    19901 (Brightspeed) & 744 & & 2050 
    \\
    5483 (Magyar Telekom) & 578 & & 1724 
    \\
		\bottomrule
	\end{tabular}
	\label{tab:as_share_tfwds_shielded}
\end{table}

\autoref{fig:no_tfwd} shows the number of transparent forwarders observed over the last year (see \autoref{sec:measurement-setup-details} for details on the measurement method and setup).
We can see that the number of transparent forwarders decreased from 620k to 530k, a reduction of 14.5\% .
Nevertheless the amount of forwarders still remains unsettlingly high.

Looking at our historic data, we notice that over the course of the last six months, from July to December 2024, around $\sim$60k (12\%) of all transparent forwarders redirect to about 25k to 30k shielded resolvers.
These shielded resolvers can be found in around 1900 different ASes, while more than half of them are part of only three ASes, \cf \autoref{tab:as_share_tfwds_shielded}.
The surface for exploitation could therefore be greatly reduced by ensuring proper configuration in these ASes alone. Furthermore we find shielded resolvers in about 130 ASes, which are only detectable through the use of transparent forwarders, with a total of about 600 transparent forwarders and 200 shielded resolvers.

\subsection{Detailed Look at Response Rate Limits}
Shielded resolvers differ in terms of deployment from public recursive resolvers.
They are protected but reachable via transparent forwarders.
To verify whether they are less or more constrained by rate limiting than public resolvers, we apply our rate limit probing approach (see \autoref{sec:experiment-setup}) to publicly accessible recursive resolvers.
We approach rate limits by increasing the transmit rate every two seconds and stopping the measurement if we receive a reply to less than 50\% of packets we sent.

\begin{figure}
	\begin{center}
		\includegraphics[width=1\columnwidth]{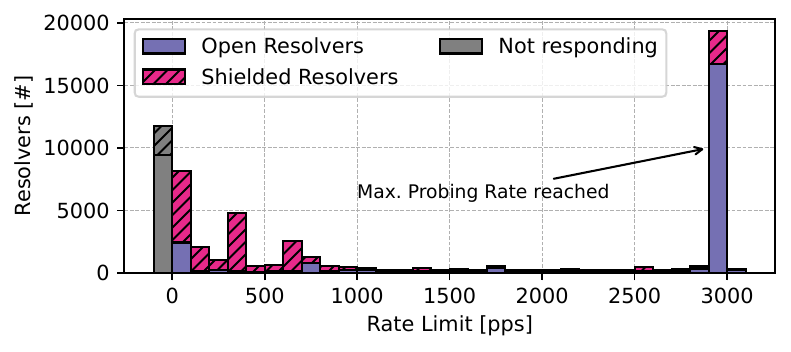}
		\caption{Response rate limits of shielded and open recursive resolvers on August 24, 2024.}
		\label{fig:resolver_ratelimits}
	\end{center}
\end{figure}

The distribution of rate limits of both shielded and open resolvers are visualized in \autoref{fig:resolver_ratelimits}.
Each bin covers a range of 100~packets per second (Pkts/s).
There are almost 2800~shielded resolvers that show no further request limit and achieve packet rates of more than 2900~packets per second.
Additionally, we observe about 17k open recursive resolvers without any measurable restriction.
These numbers are worrying as it indicates that there is no direct rate limit in place, which would restrict an attacker.

\begin{figure}
	\centering
	\includegraphics[width=\linewidth]{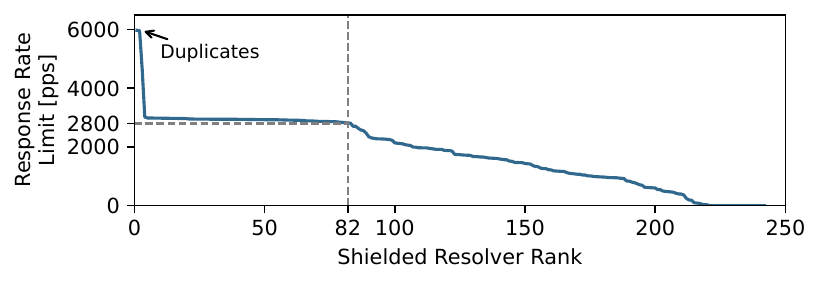}
	\caption{Response rate limits of shielded resolvers in ASes only reachable through transparent forwarders (January~6,~2025). 82 shielded resolvers are not protected by a reasonable rate limit.}
	\label{fig:rrl_new_ases}
\end{figure}

Each of the observable peaks at 0-100, 300-400, 600-700, and 2900-3000 Pkts/s are dominated by few ISPs, the same that make up more than 50\% of all transparent forwarders to shielded resolvers before, Videotron (AS5769), China Telecom (AS4812), and CenturyLink (AS209), respectively.
A more detailed look at ASes that host recursive resolvers only reachable through transparent forwarders reveals that $\approx$33\% of those resolvers are not protected by a rate limit, see \autoref{fig:rrl_new_ases}.
These resolvers directly add to the attack surface of the open DNS as they are not reachable otherwise via the global Internet, enabled by transparent forwarders.

By selecting the right transparent forwarders that expose shielded resolvers, we argue it is possible to create a substantial amount of traffic.
We can give a lower bound for achievable attack traffic as follows.
Our calculation is based on the same assumptions as discussed in \autoref{sec:tfwd_vs_rfwd} regarding the DNS query size of 70 Byte and a conservative amplification factor of 40$\times$.
The lower bound is then 180~Gbit/s.
$$\mbox{Min. traffic} = 2900\frac{Pkts}{s} \cdot 70\frac{B}{Pkt} \cdot 40 \cdot 2800 \cdot 8\frac{b}{B} \approx 180\frac{Gb}{s}$$

\subsection{Stability of Shielded Resolver Deployment}
\begin{figure}
	\begin{center}
		\includegraphics[width=1\columnwidth]{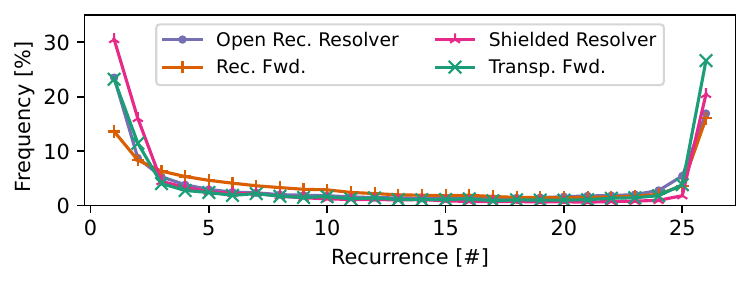}
		\caption{Recurrence of /24 subnets during weekly scans from July to December 2024.}
		\label{fig:stability_odns}
	\end{center}
\end{figure}
\autoref{fig:stability_odns} shows the recurence of /24 subnets during a six month time window, separated by ODNS response type as defined by \cite{schomp2013client, nawrocki2021transparent} adding shielded resolvers.
The data can be used as an indicator for stability of each of these components.
We can observe a U-shaped distribution, showing a large number of subnets to appear only once or twice during our scans, but also a higher percentage of subnets that can be seen consistently throughout all scans.
We expect a peak along the upper end as these are DNS resolvers and forwarders, which were purposefully set up devices as part of a stable DNS infrastructure.
On the lower end of one up to a few recurrences, we anticipate a peak as well, because these are most likely temporary setups for testing purposes, misconfigurations and consumer routing devices with low IP lease time.
This data is consistent with previous work indicating high IP churn for ODNS servers\cite{kuhrer2014exit, kuhrer2015resolvers, yazdani2022matter}.
IP churn seems to play a vital role on the selection of resolvers for misuse, as also stated by \cite{yrs-gmror-24}.
It can be observed that transparent forwarders are, with 25\% of subnets appearing across all analyzed scans, more stable than any other monitored ODNS component, while shielded resolvers appear slightly more stable than recursive forwarders and recursive resolvers. 
This means that an attacker who is aware of the dynamic ODNS landscape with high churn, can more effectively exploit the combination of transparent forwarder to shielded resolver than a recursive forwarder to a recursive resolver.

\subsection{Support of Amplification by DNSSEC and ANY Queries}
DNSSEC and ANY queries are commonly used to trigger large responses in amplification attacks.
We find that in general open recursive resolvers provide better support for \texttt{ANY}-type and DNSSEC enabled queries, which can be observed in \autoref{tab:any-dnssec-support}.
Still we see about 6000 shielded resolvers supporting DNSSEC and short of 3800 to support \texttt{ANY}-type queries.
It can be further observed that when sending \texttt{ANY}-type queries, a major portion of shielded resolvers (83.3\%) stops responding.
Even though there appear to be way less \texttt{ANY}-type query supporting shielded resolvers in total, we see indication that those which respond with higher sized fragmented payloads are actually twice as many as open recursive resolvers.
\begin{table}[t]
	\centering
	\caption{Support of ANY queries and DNSSEC for shielded and open recursive resolvers. The majority (83\%) of shielded resolvers is not responsive (\textit{n/a}) to ANY queries.}
	\label{tab:any-dnssec-support}
	\begin{tabular}{lcrrrr}
		\toprule
		 & & \multicolumn{2}{c}{Shielded Resolvers} & \multicolumn{2}{c}{Open Resolvers} \\
		\cmidrule(rl){3-4}
		\cmidrule(rl){5-6}
		\multicolumn{2}{c}{Feature Support} & [\#] & [\%] & [\#] & [\%] \\
        \midrule
        \multirow{3}{*}{DNSSEC} & \cmark & 5989 & 23.62 & \num{22092} & 65.04 \\
        & \xmark & \num{15197} & 59.94 & 4630 & 13.63 \\
        & \textit{n/a} & 4168 & 16.44 & 7246 & 21.33 \\
		\dhline
		\multirow{3}{*}{ANY} & \cmark & 3791 & 14.95 & \num{24146} & 71.09\\
        & \xmark & 505 & 1.99 & 884 & 2.60 \\
        & \textit{n/a} & \num{21058} & 83.06 & 8938 & 26.31 \\
		
		\bottomrule
	\end{tabular}
\end{table}

\section{Exploiting Anycast Infrastructure}
\label{sec:exploiting-anycast-deployments}

In this section, we show that an attacker can exploit the anycast infrastructure of distributed recursive resolvers used by transparent forwarders.
We present the distributions of anycast instances of different DNS providers, to which extent they allow for scalable sending rates, and combine our observations to potentially orchestrate transparent forwarders.

\begin{table*}
	\centering
	\caption{Anycast deployments reachable via transparent forwarders. The anycast infrastructure of Google, Cloudflare, and OpenDNS is globally reachable through transparent forwarders, therefore extending the potential attack surface. Larger regional anycast services such as Yandex DNS, 114DNS, and AliDNS with at least five ingress points are also suitable for attacks.}
	\label{tab:tfwd-and-anycast-infrastructure}
	\begin{tabular}{llrrrrr}
		\toprule
		\multicolumn{2}{c}{Public Recursive Resolver} & \multicolumn{3}{c}{Transparent Forwarder} & \multicolumn{2}{c}{Recursive Resolver Anycast Infrastructure} \\
		\cmidrule(rl){1-2}
		\cmidrule(rl){3-5}
		\cmidrule(rl){6-7}
		Provider & IP Address & Countries [\#] & IP Addresses [\#] & /24 prefix [\#] & IP Addresses [\#] & /24 Prefix [\#] \\
		\midrule
		\multirow[c]{2}{*}{Google} & 8.8.8.8 & 74 & \num{341447} & 5174 & 1889 & 139 \\
		& 8.8.4.4 & 28 & \num{14115} & 259 & 537 & 53 \\
		\dhline
		\multirow[c]{2}{*}{Cloudflare} & 1.1.1.1 & 46 & \num{48313} & 953 & 216 & 151 \\
		& 1.0.0.1 & 11 & 1196 & 27 & 39 & 23 \\
		\dhline
		OpenDNS & 208.67.222.222 & 26 & \num{14464} & 227 & 224 & 42 \\
		\dhline
		SafeDNS & 195.46.39.39 & 2 & 19 & 4 & 2 & 2 \\
		\dhline
		Level3 DNS & 4.2.2.1 & 2 & 9 & 2 & 9 & 3 \\
		\dhline
		Yandex DNS & 77.88.8.8 & 1 & 725 & 17 & 119 & 5 \\
		\dhline
		114DNS & 114.114.114.114 & 1 & 70 & 26 & 5 & 3 \\
		\dhline
		AliDNS & 223.5.5.5 & 1 & 62 & 22 & 16 & 16 \\
		\dhline
		DNSPod & 119.29.29.29 & 1 & 3 & 3 & 3 & 3 \\
		\dhline
		SkyDNS & 193.58.251.251 & 1 & 3 & 1 & 1 & 1 \\
		\dhline
		CNNIC DNS & 1.2.4.8 & 1 & 2 & 1 & 2 & 1 \\
		\bottomrule
	\end{tabular}
\end{table*}

\subsection{Distribution of Transparent Forwarders}

\begin{figure}[t]
	\centering
	\includegraphics[width=\linewidth]{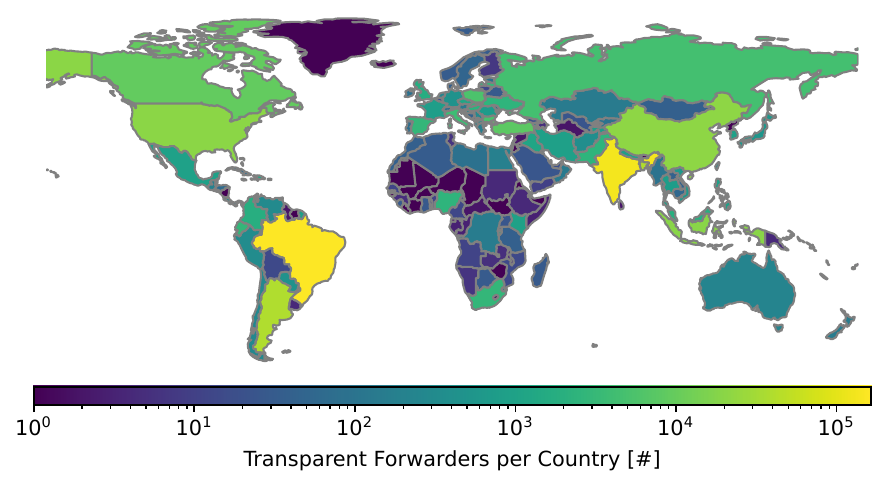}
	\caption{Overview of world-wide transparent forwarder deployment.}
	\label{fig:tfwd-deployment-overview-worldmap}
\end{figure}

\paragraph{Global distribution of transparent forwarders}
Transparent DNS forwarders are publicly accessible via the global Internet in 175~countries, with a strong bias towards Brazil~(31\%) and India (24\%), see \autoref{fig:tfwd-deployment-overview-worldmap}.
The majority of addresses are located in 2856~/24~subnets in Brazil and 2140~/24~subnets in India. 
Regarding the coverage in /24 subnet, 23\% of the potential addresses of a /24~subnet are responsive on average.
We also find 124 /24 subnets to be fully responsive in Brazil and 202 to be fully responsive in India.
In total, we find 567 /24 subnets in 26 countries to be fully responsive, representing a total of 27\% of the individual transparent forwarder IP addresses.

Our observations have two implications.
First, any attacker that (re)scans the global Internet to find vulnerable services can save scan traffic, \eg by excluding most of the addresses in fully responsive subnets.
Scanning less helps not only for scalability reasons but also to stay under the radar of preventive intrusion detection mechanisms.
Second, attackers have access to a widely distributed infrastructure.
45\% of forwarders are located in 173~countries, even though the other major part is located in only two countries.
The latter, on the other hand, gives hope to efficiently approach a smaller subset of operators to reduce the threat landscape.

\paragraph{Unveiling anycast infrastructure through transparent forwarders}
Our measurements reveal recursive resolvers that are used by transparent forwarders.
To flag recursive resolvers that are part of an IP~anycast infrastructure~\cite{RFC-1546}, we use meta data from the well-known list \url{https://www.publicdns.xyz}.

When sending DNS~queries to transparent forwarders, we are able to exploit the anycast infrastructure in 89~countries. 
The anycast infrastructure of Google alone is reachable in 79~countries via 340K transparent forwarders.
This unveils more than 2400~egress points in 192 /24 subnets. 
Similarly, we can potentially utilize the anycast infrastructure of Cloudflare and OpenDNS leveraging 479 egress points in 216 /24 subnets.
Our results are summarized in \autoref{tab:tfwd-and-anycast-infrastructure}. 

To implement an effective attack, it is crucial to select transparent forwarders carefully so that there are no overlaps between two transparent forwarders using the same anycast Point-of-Presence (PoP) at the same time to avoid rate limiting. 
PoPs closely located together are likely to share information about rate limited hosts because of aggregation effects.
Hence, attackers might tend to select transparent forwarders as far apart from each other as possible to maximize attack rates.
In the next subsection, we analyze whether this is a doable approach or whether the anycast infrastructure is protected by proper rate limits.
 
\subsection{Rate Limits of Anycast Deployments}
\begin{figure}[b]
	\begin{center}
		\includegraphics[width=1\columnwidth]{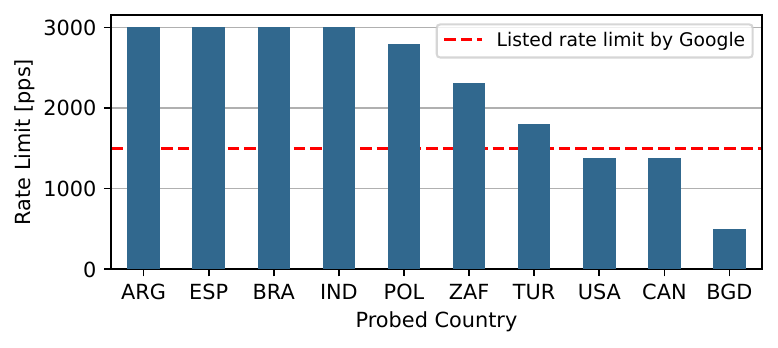}
		\caption{Regional differences of the recursive resolver rate limiting, as measured based on transparent forwarders that use the public Google resolver 8.8.8.8 (October 22, 2024).}
		\label{fig:google_different_rrl}
	\end{center}
\end{figure}

We first look at the response rate limits of popular recursive DNS resolvers from three different vantage points (Brazil, Germany, and India). 
\autoref{tab:rrl_pops} shows the average maximum response rate in packets per second from our measurement~campaign. 

Rate limits vary across DNS providers and between locations for the same provider. Some providers enforce stricter rate limits by default, which seems to be the case, \eg with Cloudflare and OpenDNS, while others limit requests based on the load of their infrastructure at the location of the PoP, \eg Google. 
We chose the public Google resolver as an example to illustrate these regional differences in \autoref{fig:google_different_rrl}.  Here, we use transparent forwarders in different countries to reach anycast instances of the same service endpoint (\ie 8.8.8.8). 
We observe a rate limit of around 1500~Pkts/s for North America, while the limit is significantly less strict in other regions such as Europe allowing at least 3000~Pkts/s for some locations. 
Interestingly, the rate limits are not necessarily stable over time.
We notice differences in the same country, \eg in India and Brazil, when conducting the measurement at another day.
Which, again, might be due load-adaptive behavior of the anycast infrastructure or reaching another PoP because of simple round~robin.

In any case, misusing transparent forwarders allows an attacker to reach multiple anycast infrastructures that deploy less strict rate limiting, which increases the threat potential of transparent forwarders significantly.

\begin{table}[t]
	\caption{Response rates at public DNS resolvers measured from vantage points in Brazil, Germany, and India in November 2024. Measurements were performed on four consecutive days, the values shown are the averages from these four days. Rate limits often depend on the vantage point.}
	\label{tab:rrl_pops}
	\newcolumntype{Y}{>{\raggedleft\arraybackslash}X}
	\begin{tabularx}{\columnwidth}{llYYY} %
		\toprule
		& & \multicolumn{3}{c}{Average Response Rate} 
		\\
		 \multicolumn{2}{c}{Public Recursive Resolver}  & \multicolumn{3}{c}{[pps]} 
		\\
		\cmidrule(lr){3-5}
		\cmidrule(lr){1-2}
		Provider & Anycast IP Address & BRA & GER & IND \\
		\midrule
		\multirow[c]{2}{*}{Google} & 8.8.4.4 & 2177 & 1773 & 1597 \\
		& 8.8.8.8 & 2170 & 1744 & 1608 \\
		\dhline
		\multirow[c]{2}{*}{Cloudflare} & 1.0.0.1 & 1008 & 874 & 754 \\
		& 1.1.1.1 & 863 & 822 & 796 \\
		\dhline
		\multirow[c]{2}{*}{OpenDNS} & 208.67.220.220 & 832 & 829 & 769 \\
		& 208.67.222.222 & 785 & 733 & 773 \\
		\dhline
		\multirow[c]{2}{*}{Verisign} & 64.6.64.6 & 192 & 199 & 192 \\
		& 64.6.65.6 & 2254 & 1796 & 1401 \\
		\dhline
		\multirow[c]{2}{*}{AliDNS} & 223.5.5.5 & 1894 & 1753 & 1341 \\
		& 223.6.6.6 & 1882 & 1729 & 1506 \\
		\dhline
		\multirow[c]{2}{*}{CNNIC DNS} & 1.2.4.8 & 1932 & 426 & 740 \\
		& 210.2.4.8 & 1194 & 1358 & 223 \\
		\dhline
		\multirow[c]{2}{*}{114DNS} & 114.114.114.114 & 58 & 55 & 54 \\
		& 114.114.115.115 & 58 & 54 & 51 \\
		\dhline
		\multirow[c]{2}{*}{DNSPod} & 119.28.28.28 & 1877 & 1831 & 1538 \\
		& 119.29.29.29 & 1966 & 1870 & 1502 \\
		\dhline
		\multirow[c]{2}{*}{Yandex DNS} & 77.88.8.1 & 25 & 28 & 26 \\
		& 77.88.8.8 & 21 & 28 & 24 \\
		\bottomrule
	\end{tabularx}
\end{table}

\subsection{Orchestration of Transparent Forwarders}

\begin{table*}
	\centering
	\caption{Packet rates achieved when orchestrating max. 10 transparent forwarders that use Google, Cloudflare, or OpenDNS as backend resolvers. An attacker requires a single host under control to launch the attack. Achieving the same attack volume by misusing recursive resolvers directly requires additional infrastructure under the control of an attacker (given a single recursive resolver that is based in Germany and packet rates per resolver as presented in \autoref{tab:rrl_pops}).}
	\label{tab:anycast_sum_rates}
	\begin{tabular}{llrrrrrr}
		\toprule
		 \multicolumn{2}{c}{Recursive Resolver Backend} & \multicolumn{3}{c}{Transparent Forwarders} &  \multicolumn{3}{c}{Recursive Resolvers} \\
		 \cmidrule(rl){1-2}
		 \cmidrule(rl){3-5}
		 \cmidrule(rl){6-8}
		\makecell[l]{\\ \\ Provider} & \makecell[l]{\\ \\ Anycast IP Address} & \makecell[r]{Attack Hosts \\ Under Control \\ \ [\#]} & \makecell[r]{Forwarders per \\ Attack Host \\ \ [\#]} & \makecell[r]{Attack \\ Volume \\ \ [pps]} & \makecell[r]{Attack Hosts \\ Under Control \\ \ [\#]} & \makecell[r]{Resolvers per \\ Attack Host \\ \ [\#]} & \makecell[r]{Attack \\ Volume \\ \ [pps]}\\
		\midrule
		\multirow{2}{*}{Google} & 8.8.8.8  & 1 & 10 &  \num{24815} & \num{14}  & 1 & \num{24416}\\
		& 8.8.4.4           & 1 & 7 &  \num{17153} & \num{10} & 1 & \num{17330} \\
		\dhline
		\multirow{2}{*}{Cloudflare} & 1.1.1.1 & 1 &  8  & \num{6002} & \num{7} & 1 & \num{6118} \\
		& 1.0.0.1           & 1 &  3 &  \num{3600} & \num{4} & 1 & \num{3288} \\
		\dhline
		\multirow{2}{*}{OpenDNS} & 208.67.222.222 & 1 &  7 &  \num{5660} & \num{8} & 1 & \num{5864} \\
		& 208.67.220.220    & 1 &  5 &  \num{3950} & \num{5} & 1 & \num{4145} \\
		\bottomrule
	\end{tabular}
\end{table*}

An attacker could orchestrate multiple transparent DNS forwarders to avoid local rate limits.
We empirically assess the packet rate that can be achieved by such orchestration.
Due to ethical concerns, we only orchestrate a maximum of 10~transparent forwarders at the same time.
To prevent traffic peaks in smaller regions, we select one transparent forwarder per country.
These measurements consider three groups of forwarders, where each group consists of forwarders using a different public resolver infrastructure.

Our measurements show that the orchestration of multiple transparent forwarders is successful.
We can achieve packet rates of nearly 25k packets by using only ten~forwarders.
These forwarders send the DNS~queries to different anycast instances of each DNS provider---all of this initiated by a single host under control of the attacker.

Alternatively, an attacker could misuse recursive resolvers directly.
For this to be successful, however, the attacker needs to control multiple hosts in different regions.
Otherwise, the attacker would only be able to target the same anycast instance, which then would limit the attack by the local rate limit.
For example, given the packet rate measured for a recursive resolver instance in Germany (see \autoref{tab:rrl_pops}) and assuming this rate available in other regions, an attacker would need 4$\times$ to 14$\times$ more infrastructure than when orchestrating transparent forwarders.
Our results are summarized in \autoref{tab:anycast_sum_rates}.

\section{Discussion}
\label{sec:discussion}

\paragraphNoDot{Can an attacker rely on a stable transparent forwarder infrastructure?}
Yes.
Transparent DNS forwarders contribute a significant part to the global open DNS infrastructure since years.
We found that transparent forwarders tend to be more stable in terms of IP~addresses than recursive forwarders.
To keep the list of potential forwarders up-to-date, an attacker can scan the full IPv4~address space in less than 45~minutes~\cite{durumeric2013zmap}, requiring few resources. 
Considering our observation of fully responsive prefixes, an attacker could reduce the amount of scan traffic further.

\paragraphNoDot{Are transparent forwarders just routers with a DNAT misconfiguration?}
A Destination Network Address Translation (DNAT) rule can cause a router to act as a transparent forwarder.
This is, however, of less importance for two reasons.
First, we can confirm that devices we fingerprinted have no ability to set such a rule manually, implying that devices are shipped with an open threat potential by default.
Second, the threat potential analyzed in this paper is independent on how the transparent forwarding behavior is achieved.

\paragraph{Over- or underestimation of the threat landscape}
Rate limiting is a well-known countermeasure.
During our measurements, we only experienced local rate limits, which are easy to avoid through the orchestration of transparent forwarders.
We know at least one large DNS provider that claims to protect its a anycast infrastructure by implementing a global rate limiting.
We were not able to verify this because of ethics considerations.
Similarly, the maximum packet rates that we observed when exploring local rate limiting should be considered a lower bound.

\paragraph{Manipulated responses may introduce measurement failures}
Any DNS measurement is prone to failures because of manipulated responses.
The Great-Firewall of China (GFW) is well known for manipulating DNS responses on path by intercepting queries to arbitrary hosts that surpass the GFW instead of forwarding a request to the authoritative server~\cite{anonymous2014greatfirewall,farnan2016poisoning}.
These manipulated responses inflate results as a scanner would incorrectly assume a running DNS service under the targeted IP address but the DNS replies do not reflect the real amplification of the requested name.
We carefully considered this case.
We run an authoritative nameserver for our zone that sends out custom responses, each with two \texttt{A} records: \one the IP address of the device that queried our nameserver, and \two a static IP address as a control record which we maintain.
Our post-processing then detects manipulations and removes them.
These DNS manipulations are different from fully responsive prefix deployments, where a DNS~resolver answers with the correct resource~records.

\paragraph{Other anomalies triggered by transparent forwarders}
Transparent forwarders may trigger uncommon behavior.
This includes mirroring DNS~replies triggered by other resolvers towards the victim, amplifying the attack traffic significantly more.
Those incidents have been observed in the past~\cite{ynhkj-hedrs-23}, which we still can confirm.

\paragraph{Mitigation options}
To mitigate attacks discussed in this paper, we suggest that network operators implement one of the following three action items.
\one Implementation of network ingress filtering~\cite{RFC-2827}.
This prevents spoofed traffic from passing through the network, hence rendering the misuse of transparent forwarders impossible.
\two Implementation of reverse path forwarding checks~\cite{RFC-8704}, which has the same effect as ingress filtering.
\three Verifying and updating firewall rules.
Sometimes, external traffic to DNS resolver infrastructure is only filtered at network borders.
In these cases, transparent forwarders enable attackers to bypass the firewall by accessing shielded resolvers.
Deploying the firewall rules at the resolver prevents unauthorized access.

\paragraph{Detection of malicious DNS traffic}
We analyzed detection rules for Zeek and Snort, two popular open-source network intrusion detection systems.
There are no ready-to-use DNS rules provided by default.
Adiwal~\etal~\cite{ars-didss-23} suggest a threshold of a 10k packets/s persisting for ten~seconds to detect DNS~floods, independent of the query name but linked to the source IP~address.
For Zeek, an external module\footnote{\url{https://github.com/jbaggs/anomalous-dns}} suggests to raise alerts when 12~packets within 45~seconds occur via the same five tuple (source and destination IP~addresses, protocol, source and destination ports).
These thresholds could capture attacks via transparent forwarders if attackers do not vary source ports. 
On the other hand, these heuristics could cause collateral damage, for example by filtering benign DNS traffic.
There is a common hesitance to deploy such filters.
MacFarland~\etal~\cite{msk-tbbft-17} found that only 10\% of all authoritative nameservers deploy rate limiting.
When deploying rate limits, global (\ie source-independent) rate limits should be used, making the decision on proper thresholds even more challenging.
However, considering a denial of service attack only from a single source ignores distributed DoS attacks, which are common in today's Internet and particularly easy to implement by orchestrating transparent forwarders.

\paragraphNoDot{Is there hope?}
Yes.
Our responsible disclosure helped to fix misconfigurations of firewall rules that exposed recursive resolvers through transparent forwarders.
Continuing measuring the transparent forwarder infrastructure is a prerequisite to identify affected networks and contact responsible stakeholders.

\vspace{-0.20cm}
\section{Conclusion}
\label{sec:conclusion}
In this paper, we systematically analyzed the attack potentials of transparent DNS forwarders. 
Our results close a gap in research, which has so far focused on identifying those components.
We found that transparent DNS forwarders significantly extend the attack surface of the open DNS infrastructure and scale up reflective amplification attacks. 
They scale better in terms of potential attack volume and provide faster response times in the majority of countries where they are located. We bridged access to shielded recursive resolvers, exposing a further attack surface of the global DNS  infrastructure. We showed  by orchestration that transparent DNS forwarders can easily bypass rate limiting of large public DNS providers. For DNS providers that implement IP anycast, we identified an additional amplification scale, which we empirically verified up to a factor of 14.
Finally, we succeeded with our responsible disclosure, removing 280k transparent DNS forwarders from the attack landscape. This reduced the attack potential by more than~30\%.

\begin{acks}
We would like to thank \'Italo Cunha and John Kristoff for providing us resources to run our distributed measurements.
This work was partly supported by the \grantsponsor{BMFTR}{Federal Ministry of Research, Technology and Space (BMFTR)}{https://www.bmbf.de/} within the projects IPv6Explorer (\grantnum{BMFTR}{16KIS1815}) and AI.Auto-Immune (\grantnum{BMFTR}{16KIS2333} and \grantnum{BMFTR}{16KIS2332K}).
\end{acks}

\label{lastpage}
\balance
\bibliographystyle{ACM-Reference-Format}
\bibliography{../bib/bibliography,../bib/rfcs,../bib/own,../bib/internet}
\appendix
\section{Ethics Considerations}
\label{sec:ethics}
We presented a method and measurement results to better understand the threat potential of open transparent DNS forwarders in principle and in the wild.

\paragraph{Stakeholders}
Usually, transparent forwarders either run on customer equipment (home gateways) or data center routers.
Our results are of relevance for network operators as they control access to their network and decide on protection mechanisms.

Furthermore, the results are relevant for any group that measures the open DNS infrastructure to inform about potential threat vectors, \eg Shadowserver.
Most of the popular on-going DNS~measurement campaigns ignore transparent forwarders and thus underestimate the threat potential.

\paragraph{Potential risks and mitigation options}
The goal of this paper is to raise awareness and to motivate the operator community to close this threat vector.
Our results, however, could also be misused to educate attackers.
We argue that our decision outweighs potential negative effects for the following reasons.
First, we conduct our measurements since more than two years.
We implemented a responsible disclosure policy and still try to approach relevant stakeholders (see below).
The level of improvement (\ie closing transparent forwarders) saturates.
Instead of continuing being silent about the details of this threat vector, we favor educating those who can help to improve the situation.
Furthermore, we believe that this paper provides structural network security insights independent of DNS.
The case of shielded resolvers, for example, does also hold for other services---motivating the verification of block lists in general.
Finally, we presented mitigation options in \autoref{sec:discussion}.

\paragraph{Responsible disclosure}
We are in contact with network operators and federal security offices to raise awareness of transparent forwarders to prevent unintentional access to those ODNS~components.
Our research results already helped to identify misconfigurations of firewalls rules that exposed shielded resolvers via transparent forwarders.
One major European~ISP fixed its deployment based on the insights presented in this paper.
We discussed the implications of our measurements with the DNS team of the European~ISP and received clearance to conduct our experiments. 
We limited the duration of these experiments to two seconds per experiment. No negative impact was reported.
We provided mitigation options, which they implemented.

\section{Artifacts}
\label{sec:artifacts}

All artifacts of this paper are publicly available.
These include
  \one all raw measurement data that we used to identify different types of open DNS components;
  \two all data derived from post-processing, \ie data that substantiate our arguments and serve as input for our figures;
  \three measurement scripts;
  \four post-processing scripts.
All artifacts and details on how to use them are archived on \url{https://doi.org/10.5281/zenodo.16998590}.

\end{document}